\documentclass[12pt]{article}
\normalbaselines
\begin{document}

\begin{center}

{\bf EXTENDED EINSTEIN-MAXWELL MODEL}

\vspace{8mm}
{\bf A.B. Balakin}

\vspace{8mm}

{\it Department  of General Relativity and Gravitation,\\
Kazan State University, Kremlevskaya str. 18, Kazan 420008,
Russia

E-mail: Alexander.Balakin@ksu.ru
}
\end{center}

\vspace{8mm}

\begin{abstract}
A self-consistent extended Einstein-Maxwell model for
relativistic non-stationary pola\-rizable-magnetizable anisotropic
media is presented. Based on the analogy with relativistic
extended irreversible (transient) thermodynamics, the extended
constitutive equations for the electrodynamics of continua are
formulated phenomenologically, the convective derivatives of the
first, second, etc. orders being taken into account. The master
equations for the gravity field contain a modified effective
(symmetric) stress-energy tensor of the electromagnetic field in a
material medium, the use of this tensor being motivated both by
historical analogies and direct variational procedure. By way of
example we consider the exact solution of the extended
Einstein-Maxwell model, describing the isotropic cosmological
model with hidden non-vanishing electromagnetic field, electric
polarization and magnetization.
\end{abstract}

\vspace{10mm}
\noindent
{\bf Key words}: {\it anisotropic medium, polarization, magnetization, Einstein-Maxwell
theory, \\
extended thermodynamics, extended constitutive equations}.

\vspace{0.8cm}


\newpage

\section{Introduction}

Non-stationarity of electromagnetically active media is known to
lead to complicated non-equilibrium phenomena, and one of the
cooperative degrees of freedom, excited by the medium dynamics,
relates to the phenomena of polarization and magnetization
\cite{EM}-\cite{Chernyi}.
The Universe is a
non-stationary system that contains numerous electrodynamic
subsystems of different scales \cite{MF1,MF2,MF3}. It is natural
to expect that the cosmological non-steady background stimulates
various evolutionary processes, involving the phenomena of
polarization and magnetization dynamics. This circumstance
attracts our attention to the exact non-stationary solutions to
the self-consistent Einstein-Maxwell equations, describing the
evolution of gravitating polarizable-magnetizable media. Following
basic Einstein's ideas, the stress and energy of polarization and
magnetization themselves act as sources of the gravity field.
These sources can not be negligible in comparison with the
contribution of pure electromagnetic field: for instance, the
plasma susceptibility parameters can be at least of the same order
than one (or greater), i.e., the contributions of the electric
polarization and pure electric field to the electric induction of
the plasma are comparable. Thus, in order to formulate a
self-consistent Einstein-Maxwell model one needs to find the
adequate energy-momentum tensor of the electromagnetically active
medium taking into account the polarization and magnetization of
the medium.

In many cosmological models the total stress-energy tensor is
presented as a sum of the stress-energy of the fluid (perfect or
viscous) and of the energy-momentum tensor of pure electromagnetic
field \cite{ExactSolutions,dina}. To take into account the
interacting (cross) terms, which include the polarization and
magnetization of the medium, one needs to overcome the so-called
``Minkowski-Abraham controversy", which followed the famous papers
of Min\-kowski \cite{Minkowski}, Einstein and Laub \cite{Einstein}
and Abraham \cite{Abraham}. In the second half of the last century
the problem of representation of the stress-energy tensor for the
electromagnetically active media revived. The modified versions of
the stress-energy tensor, related to the specific properties of
the so-called ponderomotive force, have been presented and
motivated by de Groot and Mazur \cite{Groot}, Grot \cite{Grot70},
Israel \cite{IsraelT}, Maugin \cite{Maugin78} and others (see,
Refs. \cite{Skobeltsyn}-\cite{Brevik1}, \cite{Tsypkin,Chernyi} for
reviews). Some new aspects of the interrelation between the
Minkowski and the Abraham versions of the electromagnetic
stress-energy tensors and their modifications, can be found in
\cite{Antoci1}-\cite{Garrison}. However, the solution of the
``Abraham-Minkowski controversy" might lay in a modification of
terminology. Particularly, Gordon \cite{Gordon73} introduced the
label ``pseudomomentum" in application to the Minkowski momentum.
Nelson  motivated the use of the term ``wave momentum" for the sum
of the Abraham momentum and of the Minkowski one \cite{Nelson}.
Garrison and Chiao \cite{Garrison} used the terms ``canonical and
kinetic form" of electromagnetic momentum. It is clear now, that
there are two different aspects in the problem under discussion.
The first  is connected with the correct definition of the flux
four-vector of the electromagnetic field in the material medium as
a part of the so-called {\it electromagnetic energy - momentum
tensor} \cite{Maugin78}, which appears in balance equations. The
structure of this tensor can be verified in the laboratory, and a
few experiments have been proposed for this purpose (see, e.g.,
the review \cite{Brevik1}). The second aspect is connected with
the correct construction of the so-called  {\it effective
stress-energy tensor of the electromagnetic field} as an adequate
part of the total stress-energy tensor, which appears as the
source of the gravity field in the self-consistent
Einstein-Maxwell model. It is interesting to clarify the structure
of this tensor for the cosmological applications and to check it
using the observational data. Here we focus on the problem of
representation of the second quantity, namely, on the effective
stress-energy tensor of the electromagnetic field in a polarizable
and magnetizable medium.

The paper is organized as follows. In the Section 2 we discuss the
reconstruction of the electromagnetic energy-momentum tensor on
the base of balance equations, which are the consequences of the
Maxwell equations; as well, we concern some historical aspects,
which are necessary for further consideration. In the Section 3 we
introduce a new version of the effective stress-energy tensor. For
this purpose, using Lagrange formalism, we derive directly  this
tensor for three well-known cases: vacuum, spatially isotropic
medium and uniaxial anisotropic medium. Then, based on the
obtained exact formulas, we propose an ansatz about the structure
of the effective stress-energy tensor for general case and compare
it with well-known tensors of this type. In the Section 4 we
discuss the stationary and non-stationary phenomenological
constitutive equations, linking the polarization - magnetization
tensor and its derivatives with the Maxwell tensor and its
derivatives. Thus, we introduce the so-called {\it extended}
constitutive equations for relativistic electrodynamics of
continuous media, and in this sense we get an {\it extended
Einstein-Maxwell model}. In the Section 5 we discuss the example
of the exact solution of the Einstein-Maxwell equations, based on
the proposed effective stress-energy tensor of the electromagnetic
field. To demonstrate the novelty of our approach, we discuss a
particular example, when electric and magnetic fields,
polarization and magnetization, as well as electric and magnetic
inductions in the medium are non-vanishing, nevertheless, their
cooperative contribution to the total stress-energy tensor is
equal to zero. In this exact model the Einstein equations coincide
with the ones from Friedmann-Lema\^itre-Robertson-Walker (FLRW)
model, the electromagnetic field and electromagnetic induction
being hidden from the point of view of space-time evolution.
Section 6 summarises our findings.

\section{Electromagnetic energy-momentum tensor}

\subsection{Balance equations}

The standard phenomenological way to introduce  the
electromagnetic energy - momentum tensor in a material medium is
connected with balance equations (see, e.g., \cite{Moller} for
details). The balance equations can be derived using the Maxwell
equations
\begin{equation}
\nabla_k H^{ik} = - \frac{4 \pi}{c} I^i \,, \label{0maxwEQ}
\end{equation}
\begin{equation}
\nabla_i F_{kl} + \nabla_l F_{ik} + \nabla_k F_{li} =0 \,,
\label{10maxwEQ}
\end{equation}
where $F^{ik}$ is the Maxwell tensor and $H^{ik}$ is an induction
tensor \cite{EM,LLP,HehlObukhov}. The convolution of equation
(\ref{0maxwEQ}) with $F^l_{ \ i}$
\begin{equation}
F^l_{ \ i} \nabla_k H^{ik} = - \frac{4 \pi}{c} I^i F^l_{ \ i}
\label{pondero1}
\end{equation}
can be transformed, using (\ref{10maxwEQ}), into equations, which
have an explicit  divergence form
\begin{equation}
\nabla_k {\cal T}^{kl} = F^l \,. \label{pondero2}
\end{equation}
The structure of (\ref{pondero2}) allows one to indicate the pair
${\cal T}^{kl}$ and $F^l$ as a conjugated one, ${\cal T}^{kl}$
being an {\it electromagnetic energy-momentum tensor}, and $F^l$
being a {\it ponderomotive force}. The choice for ${\cal T}^{kl}$
and $F^l$ is not unique. To fix this pair one should use some
supplementary motivation (microscopic or phenomenological).
Consider now six well-known instances.

\subsubsection{Vacuum case}

In vacuum ($I^i = 0$, $H^{ik} = F^{ik}$) equation (\ref{pondero2})
takes the conservation law form
\begin{equation}
\nabla_k {\cal T}^{kl}_{(0)} = 0  \,, \label{Tvacconserva}
\end{equation}
where
\begin{equation}
{\cal T}^{kl}_{(0)} \equiv \frac{1}{4} g^{kl}  F_{mn} F^{mn} - F^{km}
F^l_{ \ m} \label{Tvacuum}
\end{equation}
is the stress-energy tensor of the electromagnetic field. It is
symmetric, traceless, conserved and does not depend explicitly on
the velocity four-vector $U^i$ (of an observer or of the medium as
a whole). The ponderomotive force in vacuum vanishes.

\subsubsection{Minkowski version}

By rearranging equation (\ref{pondero1}) as
\begin{equation}
\nabla_k \left[
\frac{1}{4} g^{kl} H_{mn} F^{mn} - H^{km} F_{lm} \right] =
\frac{4 \pi}{c} I_i F^{il} + \frac{1}{4} \left[ F_{mn} \nabla^l
H^{mn} - H_{mn} \nabla^l F^{mn} \right] \,, \label{Poynting}
\end{equation}
we obtain the conjugated pair ${\cal T}^{kl}$ and $F^l$
\begin{equation}
{\cal T}^{kl}_{({\rm Minkowski})} \equiv  \frac{1}{4} g^{kl}
H_{mn} F^{mn} - H^{km} F^l_{ \ m} \,, \label{Minkowski1}
\end{equation}
\begin{equation}
F^{l}_{({\rm Minkowski})} = \frac{4 \pi}{c} I_i F^{il} +
\frac{1}{4} \left[ F_{mn} \nabla^l M^{mn} - M_{mn} \nabla^l
F^{mn} \right] \,. \label{Minkowski2}
\end{equation}
Here $M^{ik} \equiv H^{ik} - F^{ik}$ is the polarization -
magnetization tensor of the material medium. The tensor ${\cal
T}^{kl}_{({\rm Minkowski})}$ is traceless, but not symmetric. It
does not include explicitly the velocity four-vector of the
material medium.

\subsubsection{Modified Minkowski tensor}

Grot and Eringen \cite{GE}, Israel \cite{IsraelT}, Maugin
\cite{Maugin78}, proposed a modified version of ${\cal T}^{kl}$
and $F^l$ in the balance equation (\ref{pondero2}). These authors
discussed the following electromagnetic energy-momentum tensor and
ponderomotive force
\begin{equation}
{\cal T}^{kl}_{({\rm modif1})} \equiv
\frac{1}{4} g^{kl} F_{mn} F^{mn} - H^{km} F^l_{ \ m}  \,,
 \quad F^l = \frac{4 \pi}{c} I_m F^{ml} - \frac{1}{2} M^{mn} \nabla^l
F_{mn} \,. \label{GEMI}
\end{equation}
This tensor, whose trace does not vanish, ${\cal T}^{kl}_{({\rm
modif1})}$, differs from (\ref{Minkowski1}) in the first term.

\subsubsection{Version of Hehl and Obukhov}

Hehl and Obukhov motivated in \cite{OH} the following choice for
${\cal T}^{kl}$ and $F^l$
\begin{equation}
{\cal T}^{kl}_{({\rm modif2})} \equiv \frac{1}{4} g^{kl} F^{mn}
F_{mn} - F^{km} F^l_{ \ m} \,,
\quad
F^l = \frac{4
\pi}{c} I_m F^{ml} + F^{l}_{ \ m} \nabla_k M^{km} \,. \label{THO}
\end{equation}
The structure of this stress-energy tensor coincides  with that of
the vacuum, thus, it is symmetric, traceless and does not depend
on $U^i$. Nevertheless, in contrast to the vacuum case, it is not
a conserved quantity, and the ponderomotive force is linear in the
divergence of the polarization-magnetization tensor.

\subsubsection{Abraham's version}

Abraham \cite{Abraham} proposed to use the symmetric
electromagnetic energy-momentum tensor, which depends explicitly
on the  $U^i$, velocity four-vector of the medium as whole. When
the medium is spatially isotropic and homogeneous, the
corresponding tensor ${\cal T}^{ik}_{({\rm Abraham})}$ reads
\begin{equation}
{\cal T}^{ik}_{({\rm Abraham})} \equiv   {\cal T}^{ik}_{({\rm
Minkowski})} + (n^2-1) \Omega^i U^k \,, \label{Abraham}
\end{equation}
where $n$ is a refractive index of the medium and the vector
$\Omega^i$ is
\begin{equation}
\Omega^i = U_l (H^{il} U^m + H^{lm} U^i + H^{mi} U^l) F_{ms} U^s
\,. \label{Omega}
\end{equation}
Here and below we shall use the normalization $U^k U_k =1$. Since
$\Omega^i U_i = 0$ the tensor ${\cal T}^{ik}_{({\rm Abraham})}$ is
traceless. The corresponding ponderomotive force can be obtained
from (\ref{pondero2}).

\subsubsection{Version of de Groot and Suttorp}

De Groot and Suttorp \cite{GS}, based on a microscopic motivation,
proposed the following electromagnetic energy-momentum tensor
\begin{equation}
{\cal T}^{kl}_{({\rm modif3})} \equiv {\cal T}^{kl}_{({\rm
modif1})} - U^l U^m (F^{kn} M_{nm} - M^{kn} F_{nm}) +
U^k U^l U^m U_n F_{ms} M^{sn} \,. \label{GS}
\end{equation}
This tensor also depends explicitly on the velocity four-vector.
The corresponding ponderomotive force can be obtained from
(\ref{pondero2}).

\vspace{5mm} \noindent {\it Remark on the microscopic and
macroscopic electrodynamics.}

As it was emphasized, e.g., in \cite{GS,Antoci1}, the problem of
the choice of the electromagnetic energy-momentum tensor ${\cal
T}^{kl}$ is connected  with the basic microscopic model, as well
as, with the averaging procedure of the microscopic Maxwell
equations (the discussion about averaging procedure see, e.g., in
\cite{Zalaletdinov}). Indeed, let the microscopic electromagnetic
field $f_{ik}$ can be represented as a sum of a mean field
$F_{ik}$ and a fluctuation terms $\xi_{ik}$ with vanishing average
value $\langle \xi_{ik} \rangle = 0$. Then, one obtains, that
\begin{equation}
{\cal T}^{ik} \equiv \langle \frac{1}{4} g^{ik} f_{mn} f^{mn} -
f^{im} f^k_{ \ m} \rangle = T^{ik}_{(0)} + \langle \tau^{ik}
\rangle \,, \label{Taverage}
\end{equation}
where $T^{ik}_{(0)}$ is given by (\ref{Tvacuum}). The second term
\begin{equation}
\langle \tau^{ik} \rangle \equiv \langle \frac{1}{4} g^{ik}
\xi_{mn} \xi^{mn} - \xi^{im} \xi^k_{ \ m} \rangle
\label{Taverage1}
\end{equation}
is very sensible to the averaging procedure and essentially
depends on the microscopic model of electromagnetic interactions
in the medium.

\subsection{``DEHB" - representation of the energy-momentum tensor}

The tensor ${\cal T}^{ik}$ can be rewritten in terms of
four-vectors $D^i$, $E^i$, $H^i$ and $B^i$, which play significant
role in the covariant electrodynamics of continuous media
\cite{Maugin78}. The definitions of these four-vectors are
well-known \cite{Lichnero}
\begin{equation}
D^i \equiv H^{ik}U_k \,, \quad H_i \equiv H^{*}_{ik}U^k \,,
\quad E^i \equiv F^{ik} U_k \,, \quad B_i \equiv F^{*}_{ik} U^k \,.
\label{DHEB}
\end{equation}
$D^i$, $E^i$, $H^i$ and $B^i$ are orthogonal to the $U^i$, a
four-vector of macroscopic velocity of the medium. In its turn,
$F^{ik}$ and $H^{ik}$ can be represented as
\begin{equation}
F^{ik} = E^i U^k - E^k U^i - \eta^{ikj}B_j \,,
\quad H^{ik} = D^i U^k -  D^k U^i - \eta^{ikj}H_j \,, \label{FEBHDH}
\end{equation}
where
\begin{equation}
\eta^{ikj} \equiv \epsilon^{ikjs}U_s \,, \quad \epsilon^{ikjs}
\equiv \frac{E^{ikjs}}{\sqrt{-g}} \,. \label{eta}
\end{equation}
$\epsilon^{ikjs}$ is the Levi-Civita tensor and the term
$E^{ikjs}$ is the completely skew-symmetric Levi-Civita symbol
with $E^{0123}=1$. This tensor provides the dualization procedure:
$F^{*}_{ik} \equiv \frac{1}{2} \epsilon_{ikmn}F^{mn}$. By means of
(\ref{FEBHDH}) the tensors ${\cal T}_{({\rm Minkowski})}^{pq}$,
${\cal T}_{({\rm Abraham})}^{pq}$, ${\cal T}_{({\rm
modif1})}^{pq}$, ${\cal T}_{({\rm modif2})}^{pq}$, ${\cal
T}_{({\rm modif3})}^{pq}$ can be represented in terms of
four-vectors $D^i$, $E^i$, $H^i$ and $B^i$. For instance, the
Minkowski tensor has the form
$$
{\cal T}_{({\rm Minkowski})}^{pq} = \left(\frac{1}{2} g^{pq} - U^p
U^q \right) \left( D^m E_m + H^m B_m \right) -
$$
\begin{equation}
\left( D^p E^q {+} B^p H^q \right) {-} U^p \eta^{qmn} E_m H_n {-}
U^q \eta^{pmn} D_m B_n . \label{MinkDEBH}
\end{equation}
Taking into account the standard decomposition of this tensor
\begin{equation}
{\cal T}_{({\rm Minkowski})}^{ik} = W_{({\rm em})} U^i U^k +
U^i I^k_{(1)} + U^k I^i_{(2)} + {\cal P}_{({\rm Minkowski})}^{ik}
\,, \label{MDecomp}
\end{equation}
one can conclude that the energy density scalar $W_{({\rm em})}$,
the first and second flux four-vectors $I^i_{(1)}$, $I^i_{(2)}$
and the stress tensor ${\cal P}_{({\rm Minkowski})}^{ik}$ read,
respectively,
\begin{equation}
W_{({\rm em})} \equiv U_p {\cal T}_{({\rm Minkowski})}^{pq} U_q =
\frac{1}{2} \left( D^m E_m {+} H^m B_m \right) \,,
\label{Minkenergy}
\end{equation}
\begin{equation}
I^k_{(1)} \equiv U_p {\cal T}_{({\rm Minkowski})}^{pq} \Delta^{k}_{q} = -
\eta^k_{ \ mn} D^m B^n  \,, \label{fluxI}
\end{equation}
\begin{equation}
I^i_{(2)} \equiv \Delta^{i}_{p} {\cal T}_{({\rm Minkowski})}^{pq} U_q  =
- \eta^i_{ \ mn} E^m H^n  \,, \label{fluxII}
\end{equation}
\begin{equation}
{\cal P}_{({\rm Minkowski})}^{ik} \equiv  \Delta^{i}_{p} {\cal T}_{({\rm
Minkowski})}^{pq} \Delta^{k}_{q} =
\frac{1}{2} \Delta^{ik} \left( D^m E_m + H^m B_m \right) -
\left( D^i E^k + B^i H^k  \right) \,. \label{Minkstress}
\end{equation}
Note that the Minkowski (\ref{Minkowski1}) and Abraham (\ref{Abraham}) versions
of the energy-momentum tensor share the
same quantity $W_{({\rm em})}$. For the Abraham (symmetric)
version of the electromagnetic energy-momentum tensor, $I^i_{(1)}$
and $I^i_{(2)}$ coincide and
\begin{equation}
I^i_{(1)} = I^i_{(2)} = - \eta^i_{ \ mn} E^m H^n  \,.
\label{fluxIII}
\end{equation}
The stress tensor ${\cal P}_{({\rm Abraham})}^{ik}$ coincides with
the symmetrized one, which was obtained by Min\-kowski
\cite{Brevik1}. In terms of three-vectors $\vec{E}$, $\vec{B}$,
$\vec{D}$ and $\vec{H}$ the flux three vectors proposed by
Minkowski, read, respectively,
\begin{equation}
\vec{I}_{(1)} = [\vec{D} ,  \vec{B}] \,, \quad \vec{I}_{(2)} =
[\vec{E} ,  \vec{H}] \,, \label{fluxMA}
\end{equation}
where $[\vec{D},\vec{B}]$ denotes the vectorial product of the
three-vectors $\vec{D}$ and $\vec{B}$. In spatially isotropic
medium one has $\vec{D} = \varepsilon \vec{E}$ and $\vec{B} = \mu
\vec{H}$, where $\varepsilon$ and $\mu$ are the scalars of
electric and magnetic permeability, respectively. Thus, using the
standard definition for the Poynting flux three-vector,
$\vec{S}_{({\rm Poynting})} \equiv [\vec{E} ,  \vec{H}]$, one can
write for the corresponding three-vectors of the momentum of the
electromagnetic field
\begin{equation}
\vec{S}_{({\rm Abraham})} = \vec{S}_{({\rm Poynting})} \,,
\quad \vec{S}_{({\rm Minkowski})} = \varepsilon \mu \vec{S}_{({\rm
Poynting})} \,. \label{fluxMAP}
\end{equation}

\section{Effective stress-energy tensor}

\subsection{Lagrange formalism}

The Einstein field equations
\begin{equation}
R^{ik} - \frac{1}{2} g^{ik} R = \Lambda g^{ik} + \kappa
T^{ik}_{({\rm total})} \,, \label{Ein}
\end{equation}
must have on their right-hand side the so-called total
stress-energy tensor $T^{ik}_{({\rm total})}$, which must be
symmetric by definition and divergence-free due to the Bianchi
identities \cite{MTW}, i.e.,
\begin{equation}
T^{ik}_{({\rm total})} = T^{ki}_{({\rm total})} \,, \quad \nabla_k
T^{ik}_{({\rm total})} = 0 \,. \label{Tsymm}
\end{equation}
$R^{ik}$ is the Ricci tensor, $R$ is the Ricci scalar, associated
with metric $g_{ik}$ and $\Lambda$ is the cosmological constant.
Following the standard variation procedure one can define
$T^{ik}_{({\rm total})}$ as
\begin{equation}
T^{ik}_{({\rm total})} \equiv -
\frac{1}{\sqrt{-g}}\frac{\delta}{\delta g_{ik}} \left( \sqrt{-g}
\pounds \right) \,, \label{Teff}
\end{equation}
where the scalar $\pounds$ denotes the Lagrangian of the whole
system, and includes the terms related to the electromagnetic
field, the polarization and magnetization of the medium. The main
problem is how to separate the contribution of the pure
electromagnetic field, the contribution of the polarization and
the magnetization and the contribution of the pure matter. This
problem seems analogous to the problem of separation of pure
gravitational energy-momentum and the energy-momentum of the
medium, which is characterized by the gravitational
self-interaction. One can extract from the total stress-energy
tensor $T^{ik}_{({\rm total})}$ the {\it electromagnetic
energy-momentum tensor} ${\cal T}^{ik}$ (see, previous Section),
which, in general, is not necessarily symmetric and traceless. On
other hand, based on the variation procedure with respect to
metric, we can draw from $T^{ik}_{({\rm total})}$  the so-called
{\it effective stress-energy tensor of the electromagnetic field}
$T^{ik}_{({\rm eff})}$,  which is symmetric and traceless by
definition. Below we will distinguish between ${\cal T}^{ik}$ and
$T^{ik}_{({\rm eff})}$.
In order to motivate our ansatz about the effective stress-energy
tensor, let us, first, consider the variation procedure of its
derivation for the simplest action functional
\begin{equation}
S[F_{mn}, g_{pq}] = \int d^4 x \sqrt{-g} \left\{ \frac{R+ 2
\Lambda}{\kappa}  + L_{({\rm matter})} + \frac{1}{2} C^{ikmn} F_{ik} F_{mn}
\right\} \,. \label{lagran}
\end{equation}
Here $C^{ikmn}$ is the linear response tensor, which describes the
influence of matter to the electromagnetic field. This tensor has
the following symmetries
\begin{equation}
 C^{ikmn} = - C^{kimn} = - C^{iknm} = C^{mnik}
\,. \label{eldacoeff}
\end{equation}
Variation of $S[F_{mn}, g_{pq}]$ with respect to the four-vector
electromagnetic potential $A_i$ gives the Maxwell equations
\begin{equation}
\nabla_k H^{ik} = 0 \,, \quad H^{ik} \equiv C^{ikmn} F_{mn} \,,
\label{maxwell}
\end{equation}
where $H^{ik}$ is the induction tensor and the current four-vector
$I^i$ is absent. The variation with respect to metric tensor
$g_{pq}$ yields the Einstein equations (\ref{Ein}) with explicit
decomposition
\begin{equation}
T_{({\rm total})}^{pq} = T_{({\rm matter})}^{pq} + T_{({\rm
eff})}^{pq} \,. \label{EinTotal}
\end{equation}
As usual, the symmetric stress - energy tensor of the material
medium $T_{({\rm matter})}^{pq}$ reads
\begin{equation}
T^{ik}_{({\rm matter})} {=} W U^i U^k + q^i U^k + q^k U^i - P
\Delta^{ik} + \Pi^{ik} \,, \label{Tmatter}
\end{equation}
where $W$ is an energy density scalar of the matter, $U^i$ is a
macroscopic velocity four-vector of the medium as whole, $q^i$ is
a heat-flux four-vector, $P$ is the Pascal pressure, $\Delta^{ik}
\equiv g^{ik} - U^i U^k$ is a projector and $\Pi^{ik}$ is an
anisotropic pressure tensor. The form of the electromagnetic part
of the total stress-energy tensor, $T_{({\rm eff})}^{pq}$, depends
on the suggestions about a structure of $C^{ikmn}$ tensor. When
$C^{ikmn}$ incorporates the metric $g_{pq}$ only, the effective
stress-energy tensor takes the form
\begin{equation}
T_{({\rm eff})}^{pq} {=} \frac{1}{4} g^{pq} C^{ikmn} F_{ik} F_{mn}
- \frac{1}{2} K^{pqikmn} F_{ik} F_{mn} \,, \label{Tem}
\end{equation}
where the tensor $K^{pqikmn}$ is a formal variation derivative
\begin{equation}
K^{pqikmn} \equiv \frac{\delta}{\delta g_{pq}} C^{ikmn}  \,.
\label{Kpqikmn}
\end{equation}
$T_{({\rm eff})}^{pq}$ is, by definition, a symmetric tensor,
whose trace vanishes when
\begin{equation}
g_{pq} K^{pqikmn} = 2 C^{ikmn}  \,. \label{spurK}
\end{equation}
In this paper we assume, that the tensor $C^{ikmn}$ contains the
metric only. When $C^{ikmn}$ contains the Riemann tensor, the
Ricci tensor and the Ricci scalar, the corresponding effective
stress-energy tensor includes the covariant derivatives of the
Maxwell tensor up to the second order, and we deal with the
so-called {\it non-minimal} Einstein - Maxwell theory
\cite{BL05,BZ05}. When $C^{ikmn}$ includes the covariant
derivative of the velocity four-vector $\nabla_i U_k$, the
effective stress-energy tensor involves first covariant derivative
of $F_{mn}$ and we deal with dynamo-optical effects \cite{AlBa}.
But it is worth stressing once again that here we restrict
our-selves by the first case only, namely, when $C^{ikmn} =
C^{ikmn}[g_{pq}]$.

\subsection{Structure of $C^{ikjl}$}

When $C^{ikjl} = C^{jlik}$, a standard decomposition  of the
$C^{ikjl}$ tensor in terms of dielectric permeability tensor
$\varepsilon^{ik}$, magnetic impermeability $(\mu^{-1})_{ik}$ tensor
and magneto-electric tensor $\nu_i^{ \ k}$ is permissible
$$
C^{ikmn} = \frac12 \left[ \varepsilon^{im} U^k U^n -
\varepsilon^{in} U^k U^m + \varepsilon^{kn} U^i U^m - \varepsilon^{km} U^i U^n
\right] -
$$
\begin{equation}
\frac12 \eta^{ikl}(\mu^{-1})_{ls}  \eta^{mns} {-}
\frac12 \left[\eta^{ikl}(U^m \nu_{l}^{ \ n} {-} U^n \nu_{l }^{ \
m}) {+} \eta^{lmn}(U^i \nu_{l}^{ \ k} {-} U^k \nu_{l}^{ \ i} )
\right] \,, \label{Cdecomp}
\end{equation}
where
\begin{equation}
\varepsilon^{im} = 2 C^{ikmn} U_k U_n\ , \quad (\mu^{-1})_{pq} = -
\frac{1}{2} \eta_{pik} C^{ikmn} \eta_{mnq}\ ,
\quad \nu_{p}^{ \ m} = \eta_{pik} C^{ikmn} U_n \,. \label{emunu}
\end{equation}
The tensors $\varepsilon_{ik}$ and $(\mu^{-1})_{ik}$ are
symmetric, but $\nu_{l}^{ \ k}$ is, in general, non-symmetric.
These three tensors are orthogonal to $U^i$,
\begin{equation}
\varepsilon_{ik} U^k = 0 \,, \quad  (\mu^{-1})_{ik} U^k = 0 \,, \quad
 \nu_{l}^{ \ k} U^l = 0 = \nu_{l}^{ \ k} U_k \,.
\label{ortho}
\end{equation}
The case $C^{ikjl} \neq C^{jlik}$ is described in detail by Hehl
and Obukhov \cite{HehlObukhov}. The decomposition (\ref{Cdecomp})
leads to the well-known formulas for the four-vectors of electric
induction $D^i$, magnetic field $H_i$, electric field $E^i$ and
magnetic induction $B_i$:
\begin{equation}
D^i = \epsilon^{ik} E_k - \nu^{ \ i}_{k} B^k \,, \quad H_i = \nu^{
\ k}_{i} E_k + (\mu^{-1})_{ik} B^k \,. \label{DEBHEB}
\end{equation}
The material tensors $\varepsilon^{ik}$, $(\mu^{-1})_{ik}$,
$\nu_{l}^{\ k}$ and $C^{ikmn}$ can be decomposed using the
standard tetrad representation
\begin{equation}
{\cal S}^{i_1 i_2...i_n} = X^{i_1}_{(a_1)} X^{i_2}_{(a_2)} \cdot
... \cdot X^{i_n}_{(a_n)} {\cal S}^{(a_1)(a_2) ...(a_n)}\,.
\label{XXXX}
\end{equation}
Here the symbol $X^i_{(a)}$ denotes the set of the four tetrad
vectors, whose index $(a)$ runs over $(0),(1),(2),(3)$, and
$X^i_{(0)} \equiv U^i $. These four four-vectors are assumed to
satisfy the orthogonality - normalization rules
\begin{equation}
g_{ik} X^i_{(a)}X^k_{(b)} = \eta_{(a)(b)} \,, \label{tetra1}
\end{equation}
\begin{equation}
\eta^{(a)(b)} X^p_{(a)}X^q_{(b)} = g^{pq} \,, \label{tetra111}
\end{equation}
where $\eta_{(a)(b)}$ denotes the Minkowski matrix, diagonal
$(1,{-}1,{-}1,{-}1)$. Since the tetrad four-vectors are linked by
the relation containing the metric, for further consideration we
have to define the formula for the variation $\frac{\delta
X^i_{(a)}}{\delta g^{pq}}$.

\subsection{Variation of the tetrad vectors}

Varying the relations (\ref{tetra1}) with respect to the metric,
we obtain
\begin{equation}
X_{k(b)} \delta X^k_{(a)} + X_{k(a)} \delta X^k_{(b)}  = -
X^i_{(a)}X^k_{(b)} \delta g_{ik} \,. \label{tetra2}
\end{equation}
The variation of (\ref{tetra111}) yields
\begin{equation}
\delta g^{pq} = \eta^{(a)(b)} \left[ X^q_{(b)} \delta X^p_{(a)} +
X^p_{(a)} \delta X^q_{(b)} \right]  \,. \label{tetra3}
\end{equation}
The relations (\ref{tetra2}) and (\ref{tetra3}) are equivalent,
since
\begin{equation}
g^{ik}g_{kj} = \delta^i_j  \ \  \Rightarrow \ \ \delta g_{ik} = -
\delta g^{pq}g_{pi} g_{qk} \,. \label{tetra4}
\end{equation}
The variation of arbitrary origin $\delta X^i_{(a)}$ (not
necessarily caused by the metric variation) can be represented as
a linear combination  of the tetrad  four-vectors:
\begin{equation}
\delta X^i_{(a)} = X^i_{(c)} Y^{ \ \ (c)}_{(a)} \,. \label{tetra5}
\end{equation}
The tetrad tensor $Y^{ \ \ (c)}_{(a)}$ is not generally symmetric.
Using the convolution of (\ref{tetra3}) with tetrad vectors, we
obtain
\begin{equation}
Y^{(a)(b)} + Y^{(b)(a)} = \delta g^{pq} X_{p}^{(a)}X_{q}^{(b)} \,,
\label{tetra6}
\end{equation}
where we use the standard rules for the indices, e.g.,
$X_{q}^{(f)} = \eta^{(f)(b)} g_{qm} X^m_{(b)}$. Consequently, the
symmetric part of the quantity $Y^{(a)(b)}$, indicated as
$Z^{(a)(b)}$, can be readily found:
\begin{equation}
Z^{(a)(b)} = \frac{1}{2} \delta g^{pq} X_{p}^{(a)}X_{q}^{(b)} \,,
\label{tetra7}
\end{equation}
and the law (\ref{tetra5}) reads now
\begin{equation}
\delta X^i_{(a)} = \frac{1}{4} \delta g^{pq} \left[X_{p (a)}
\delta^i_q + X_{q (a)} \delta^i_p \right] + X^i_{(c)} {\cal
Z}_{(a)}^{\cdot (c)} \,. \label{tetra8}
\end{equation}
Here ${\cal Z}_{(a)}^{\cdot (c)}$ is the skew-symmetric part of
$Y_{(a)}^{(c)}$, i.e., $2 {\cal Z}_{(a)(c)} {=} Y_{(a)(c)} {-}
Y_{(c)(a)}$.  Therefore, the variation of the metric produces the
variation of the tetrad, described by (\ref{tetra8}) with
vanishing skew-symmetric part ${\cal Z}_{(a)(c)}$. Thus, one
finally has
\begin{equation}
\frac{\delta X^i_{(a)}}{\delta g^{pq}} = \frac{1}{4}  \left[X_{p
(a)} \delta^i_q + X_{q (a)} \delta^i_p \right] \,, \label{tetra9}
\end{equation}
and we can use this formula for the variation of the four-velocity
vector $U^i$ and for the variation of the space-like vectors
$X^i_{(\alpha)}$ ($\alpha = 1,2,3$).

Note that in \cite{Bilyalov} the author, considering the
stress-energy tensor for the spinor field, has used the formula
for the variation of tetrad, which can be easily transformed into
(\ref{tetra8}). The authors of \cite{Antoci1} - \cite{Antoci3} use
a different formula for the variation of the four-velocity vector:
\begin{equation}
\delta U^i(s) = \delta \left(\frac{dx^i}{ds}\right) =
 - \frac{1}{2} U^i \delta g_{pq} U^p U^q \,.
\label{tetra10}
\end{equation}
This formula is derived on the base of kinematic representation of
the four-velocity $U^i(s)$ as a tangent vector for the observer
world-line. Such a representation is adequate to the Lagrangian
variation procedure with respect to coordinates. However, it is
not possible to use the same method for the calculation of the
variation of the space-like vectors $X^i_{(\alpha)}$.

Our approach is based on the consideration of the four vector
fields $X^i_{(a)}(x)$, subjected to the orthogonality and
normalization  conditions. This representation is appropriate for
the  procedure of variation of the Lagrangian (\ref{lagran}) with
respect to the $A_i(x)$ and $g_{pq}(x)$ fields, as well as with
respect to spinor field, as it follows from \cite{Bilyalov}.

\subsection{Three standard examples}

\subsubsection{Pure vacuum}

The tensor $C^{ikmn}$ of vacuum must be constructed from the
metric only,
\begin{equation}
C^{ikmn} = \frac{1}{2} \left( g^{im}g^{kn} - g^{in}g^{km} \right)
\,. \label{Cvacuum}
\end{equation}
Then
\begin{equation}
K^{ \ \ ikmn}_{pq} =  \frac{1}{2} \left( \delta^i_p \delta^m_q
g^{kn} + \delta^k_p \delta^n_q g^{im} - \delta^i_p \delta^n_q g^{km} - \delta^k_p \delta^m_q
g^{ik} \right) \,, \label{Kvacuum}
\end{equation}
and the direct variation in (\ref{Kpqikmn}) gives the standard
formula for the symmetric traceless stress-energy tensor of the
electromagnetic field in vacuum (\ref{Tvacuum}).

\subsubsection{Spatially isotropic medium}

\noindent By contrast to the vacuum case the Lagrangian of
electromagnetic field in the spatially isotropic medium should
contain one supplementary vector, namely, the velocity four-vector
of the medium, $U^i$. The linear response tensor has the following
form \cite{Maugin78,Tsypkin}:
\begin{equation}
C^{ikmn} {=} \frac{1}{2\mu} \left[ \left( g^{im}g^{kn} {-}
g^{in}g^{km} \right) {+} (\varepsilon \mu {-}1) \left( g^{im} U^k U^n {-} g^{in}
U^k U^m {+} g^{kn} U^i U^m {-} g^{km} U^i U^n \right)
\right] \,, \label{Cisotr}
\end{equation}
where $\varepsilon$ and $\mu$ are the dielectric and magnetic
permeability scalars, respectively. The  $\varepsilon^{ik}$,
$(\mu^{-1})_{pq}$ and $\nu^i_{ \ k}$ tensors, entering the linear
response tensor (see, (\ref{emunu})) read
\begin{equation}
\varepsilon^{ik} = \varepsilon \Delta^{ik} \,, \quad
(\mu^{-1})_{pq} = \frac{1}{\mu} \Delta_{pq} \,, \quad \nu^i_{ \ k}
= 0 \,. \label{emunuISOTR}
\end{equation}
Since the velocity four-vector is normalized as $U_k U^k {=} 1$,
we should use the variation of $U^i$ with respect to $g_{pq}$ in
order to obtain the $K^{pqikmn}$ tensor (\ref{Kpqikmn}). To do
this  we employ the formula
\begin{equation}
\delta U^i = \frac{1}{4} \delta g^{pq} \left(U_p \delta^i_q + U_q
\delta^i_p \right) \label{deltaU}
\end{equation}
as a particular case of (\ref{tetra9}) with $(a)=(0)$. A
straightforward calculation shows that the corresponding effective
stress-energy tensor can be written as
\begin{equation}
T^{kl}_{({\rm eff})} \equiv  \frac{1}{4} g^{kl} C^{mnpq}
F_{mn}F_{pq}  - \frac{1}{2} (C^{kmpq} F^l_{ \ m} +  C^{lmpq} F^k_{ \ m} )F_{pq}
\,. \label{BPisotrop}
\end{equation}

\subsubsection{Medium with uniaxial symmetry}

\noindent The tensor $C^{ikmn}$ for the uniaxial symmetry contains
not only the velocity four-vector $U^i$, but one supplementary
space-like four-vector, (say, $X^i$) as well. This vector is
normalized according to $g_{ik}X^i X^k = -1$, and orthogonal to
$U^i$, i.e., $g_{ik} X^i U^k = 0$. Thus, to calculate the tensor
$K^{pqikmn}$ we must find the variation of $X^i$ with respect to
the metric $g_{pq}$ in addition to the variation $\delta U^i$. For
this purpose we use the formula
\begin{equation}
\delta X^i = \frac{1}{4} \delta g^{pq} \left(X_p \delta^i_q + X_q
\delta^i_p \right) \label{deltaX}
\end{equation}
as a particular case of (\ref{tetra9}). To modify the $C^{ikmn}$
tensor for the uniaxial case we follow the procedure described in
\cite{BaZiGRG}.  First, we modify (\ref{emunuISOTR}) by
\begin{equation}
\varepsilon^{ik} = \varepsilon \left( \Delta^{ik}  + \xi X^i X^k
\right) \,,
\quad (\mu^{-1})_{pq} = \frac{1}{\mu} \left( \Delta_{pq} +  \zeta X_p
X_q \right)\,, \label{emunuUNIAX}
\end{equation}
and consider the medium without magnetoelectric cross-terms, i.e.,
require the following relation to hold $\nu^{ik} = 0$. Here $\xi$
is a coefficient of anisotropy of the dielectric permeability. In
the uniaxial case it is unique, and can be defined, for instance,
as $\xi = 3 - \varepsilon^k_k / \varepsilon$. This parameter
vanishes when medium is spatially isotropic. $\zeta$ is a
corresponding coefficient of anisotropy of the magnetic
permeability. Then, we use (\ref{emunuUNIAX}) in (\ref{Cdecomp}),
and after a long but otherwise straightforward calculation we
obtain the expressions for $K^{pqikmn}$ and $T_{({\rm
eff})}^{pq}$. The result was to be expected: we recover the
formula (\ref{BPisotrop}) for the $T_{({\rm eff})}^{pq}$.

\subsubsection{General case}

In general, the tensor of linear response can be represented as
\begin{equation}
C^{ikmn} = X^i_{(a)} X^k_{(b)}X^m_{(c)}X^n_{(d)}
C^{(a)(b)(c)(d)}\,. \label{CXXXX}
\end{equation}
Our ansatz is that the tetrad quantity $C^{(a)(b)(c)(d)}$ does not
depend on the metric, and the variation with respect to $g^{pq}$
reduces to the variation of the tetrad four-vectors only, i.e., to
the formula (\ref{tetra9}). Under such an assumption using
(\ref{tetra9}) and (\ref{tetra1}) we obtain again the expression
(\ref{BPisotrop}).

\subsection{Ansatz}

Based on the direct derivation of the tensor $T^{kl}_{({\rm
eff})}$ together with the Lagrangian (\ref{lagran}), and taking
into account the common structure (\ref{BPisotrop}) for all three
well-known models with the constitutive equations $H^{ik} =
C^{ikmn} F_{mn}$, in which $C^{ikmn}$ contains the metric only, we
propose to consider the following effective  stress-energy tensor
of the electromagnetic field {\it with general constitutive
equations}
\begin{equation}
T^{kl}_{({\rm eff})} \equiv  \frac{1}{4} g^{kl} H_{mn} F^{mn} {-}
\frac{1}{2} (H^{km} F^l_{ \ m} {+}  H^{lm} F^k_{ \ m} ) \,.
\label{BPSET}
\end{equation}
If we use the effective stress-energy tensor $T^{kl}_{({\rm
eff})}$ as an electromagnetic energy - momentum tensor ${\cal
T}^{kl}$ in the relation (\ref{pondero2}), the corresponding
ponderomotive force takes the form
\begin{equation}
F_{l}^{ ({\rm eff})} = \frac{4 \pi}{c} I^i F_{il} + \frac{1}{4}
\left[ F_{mn} \nabla_l M^{mn} - M_{mn} \nabla_l F^{mn} \right] +
\frac{1}{2} \nabla_k \left[ M^{km} F_{lm} - F^{km} M_{lm}
\right] \,. \label{BPF}
\end{equation}
The effective stress-energy tensor (\ref{BPSET}) coincides with
the symmetrized Minkowski electromagnetic energy-momentum tensor
 (\ref{Minkowski1}). Thus, it is manifestly
symmetric, traceless and does not depend explicitly on the choice
of $U^i$.

\subsubsection{DEHB - representation of the effective stress-energy tensor}

In terms of four-vectors $D^i$, $E^i$, $H^i$ and $B^i$ the tensor
$T_{({\rm eff})}^{pq}$ can be represented as follows
$$
T_{({\rm eff})}^{pq} = \left(\frac{1}{2} g^{pq} - U^p U^q \right)
\left( D^m E_m + H^m B_m \right) -
$$
\begin{equation}
- \frac{1}{2} \left( D^p E^q + D^q E^p + H^p B^q + H^q B^p \right)
- \frac{1}{2} \left( U^p \eta^{qmn} + U^q \eta^{pmn} \right)
\left( D_m B_n + E_m H_n  \right) \,. \label{TDEBH}
\end{equation}
The energy density scalar $W_{({\rm eff})}$, the flux four vector
$I^i_{({\rm eff})}$ and the stress tensor ${\cal P}_{({\rm
eff})}^{ik}$ read, respectively,
\begin{equation}
 W_{({\rm eff})} \equiv U_p
T_{({\rm eff})}^{pq} U_q = - \frac{1}{2} \left( D^m E_m + H^m B_m
\right) \,, \label{energy}
\end{equation}
\begin{equation}
I^i_{({\rm eff})} \equiv  U_p T_{({\rm eff})}^{pq} \Delta^{i}_q =
\Delta^{i}_p T_{({\rm eff})}^{pq} U_q =
- \frac{1}{2} \eta^i_{\cdot mn} \left( D^m B^n + E^m H^n \right)
\,, \label{flux}
\end{equation}
\begin{equation}
{\cal P}_{({\rm eff})}^{ik} \equiv  \Delta^{i}_p T_{({\rm
eff})}^{pq} \Delta^{k}_q = \frac{1}{2} \Delta^{ik} \left( D^m
E_m {+} H^m B_m \right) {-}
\frac{1}{2} \left( D^i E^k {+} D^k E^i {+} H^i B^k {+} H^k B^i
\right) \,. \label{stress}
\end{equation}
Note that the energy density scalar $W_{({\rm eff})}$ coincides
with $W_{({\rm em})}$ given by (\ref{Minkenergy}), obtained for
the Minkowski and the Abraham tensors. The flux four-vector
$I^i_{({\rm eff})}$ is one half of the sum $I^i_{(1)}$ and
$I^i_{(2)}$ (\ref{fluxI}),(\ref{fluxII}). The stress tensor ${\cal
P}_{({\rm eff})}^{ik}$ is the symmetrized one, obtained by
Minkowski, and coincides with the Abraham stress-tensor. In
spatially isotropic medium in the three-vector symbols we have for
our definition of the effective stress-energy tensor
\begin{equation}
\vec{S}_{({\rm eff})} = \frac{1}{2} (\varepsilon \mu  + 1)
\vec{S}_{({\rm Poynting})} =
\frac{1}{2} (\varepsilon \mu  + 1) \vec{S}_{({\rm Abraham})} =
\frac{(\varepsilon \mu  + 1)}{2 \varepsilon \mu} \vec{S}_{({\rm
Minkowski})} \,. \label{fluxBP}
\end{equation}

\section{Constitutive equations}

The ansatz (\ref{BPSET}) concerning the structure of the effective
stress-energy tensor of the electromagnetic field in a material
medium allows us to consider self-consistently an Einstein-Maxwell
model for the evolution of non-stationary electromagnetically
active media. This model includes, first, the Einstein field
equations (\ref{Ein}) with (\ref{EinTotal}), (\ref{Tmatter}) and
(\ref{BPSET}), secondly, the Maxwell equations (\ref{0maxwEQ}) and
(\ref{10maxwEQ}), thirdly, the so-called constitutive equations,
linking the polarization - magnetization tensor $M^{ik} = H^{ik} -
F^{ik}$ and Maxwell tensor $F_{mn}$. The specific feature of the
interaction between a non-stationary material medium and the
electromagnetic field is the dynamics of polarization and
magnetization, and the constitutive equations must reflect the
existence of an inertia in the electromagnetic response of the
medium. In order to motivate our ansatz about the constitutive
equations, we briefly consider the well-known classical analogs.

\subsection{Classical and relativistic extended thermodynamics as a hint
for the construction of the covariant extended continuum
electrodynamics}

\subsubsection{Rheological models}

The well-known Hook law
\begin{equation}
\sigma^{\alpha \beta} = {\cal C}^{\alpha \beta \gamma \rho}
e_{\gamma \rho} \equiv \sigma^{\alpha \beta}_{({\rm
stationary})}\,, \label{hook}
\end{equation}
describing the linear relation between the stress tensor
$\sigma^{\alpha \beta}$ and the deformation tensor $e_{\gamma
\rho}$, is considered to be the simplest stationary linear
constitutive law in rheology \cite{Sedov}. Greek indices run over
$1,2,3$ and describe the spatial coordinates. The material tensor
${\cal C}^{\alpha \beta \gamma \rho}$ is symmetric, i.e.,
\begin{equation}
{\cal C}^{\alpha \beta \gamma \rho} = {\cal C}^{ \beta \alpha
\gamma \rho} = {\cal C}^{\alpha \beta \rho \gamma} = {\cal C}^{
\gamma \rho \alpha \beta} \,, \label{elacoeff}
\end{equation}
and includes the elastic coefficients  of the  medium \cite{Nye}.
When the medium is non-stationary, the difference $\sigma^{\alpha
\beta} - \sigma^{\alpha \beta}_{({\rm stationary})}$ becomes a
function of time due to inertia effects. Maxwell's viscosity model
\cite{Maxwell} assumes, that this difference is proportional to
the time derivative of the stress tensor. In the Kelvin-Voigt
model (we follow the terminology of Ref. \cite{JCL}) the
difference $\sigma^{\alpha \beta} - \sigma^{\alpha \beta}_{({\rm
stationary})}$ is proportional to the time derivative of the
deformation tensor. The Poynting-Thomson model (see, \cite{JCL})
combines Maxwell and Kelvin-Voigt models and is characterized by
the constitutive law
\begin{equation}
\sigma^{\alpha \beta} - {\cal C}^{\alpha \beta \gamma \rho}
e_{\gamma \rho} =  - \Gamma^{\alpha \beta}_{\gamma
\rho}\frac{\partial}{\partial t}\sigma^{\gamma \rho} +
\lambda^{\alpha \beta \gamma \rho} \frac{\partial}{\partial t}
e_{\gamma \rho} \,. \label{thomson}
\end{equation}
Here the first term in the right-hand-side corresponds to the
Maxwell model and the second term relates to the Kelvin-Voigt
contribution. The quantities $\Gamma^{\alpha \beta}_{\gamma \rho}$
and $\lambda^{\alpha \beta \gamma \rho}$ represent the tensors of
relaxation parameters for stresses and strains, respectively.
These three basic models provide a rule for the next
generalizations of the rheological models (see, e.g.,
\cite{Maugin99}), namely, to introduce the successive derivatives
of the stress tensor and/or strain tensor (see, e.g., Jeffreys
model \cite{JCL}), by the same method, used for the first
derivative.

\subsubsection{Heat conduction model}

The Fourier law
\begin{equation}
\vec{q} = - \lambda \vec{\nabla} T  \equiv \vec{q}_{({\rm
stationary})}  \label{fourier}
\end{equation}
is a stationary constitutive law for the heat conduction,
connecting the heat flux vector $\vec{q}$ with the spatial
gradient of the temperature. Taking into account the inertial
properties of the heat propagation,  Cattaneo \cite{Cattaneo48}
considered the difference $\vec{q} - \vec{q}_{({\rm stationary})}$
to be linear in the time derivative of the heat flux vector
\begin{equation}
\vec{q} + \lambda \vec{\nabla} T = - \tau_{(q)} \frac{\partial
{\vec{q}}}{\partial t} \,. \label{cattaneo}
\end{equation}
The latter expression, supplemented with the balance equation for
the internal energy leads to the hyperbolic equation, governing
the temperature evolution
\begin{equation}
\tau_{(q)} \frac{\partial^2}{\partial t^2} T +
\frac{\partial}{\partial t} T  = \chi \vec{\nabla}^2 T  \,,
\label{hyperT}
\end{equation}
which generalizes the standard parabolic equation. By extending of
the constitutive law for the anisotropic media \cite{Nye} and
using Cattaneo's proposal of the inertia of heat and its
extensions \cite{Jou1}, one obtains the generalized
Fourier-Cattaneo law
\begin{equation}
q^{\alpha} = - \lambda^{\alpha \beta} \nabla_{\beta} T -
Q^{\alpha}_{\beta} \frac{\partial}{\partial t} q^{\beta} +
\chi^{\alpha \beta} \nabla_{\beta} \frac{\partial T}{\partial t} +
... \,. \label{cattaneoG}
\end{equation}

\subsubsection{Relaxation of the electric polarization and magnetization}

Taking into account the delay in the response of the medium to the
applied electromagnetic field, one can use the simplest extended
constitutive equations containing the first time derivative
\cite{Groot}
\begin{equation}
\vec{P} = \chi \vec{E} - \tau_{(p)} \frac{\partial}{\partial
t}{\vec{P}} \,, \quad  \vec{M} = \chi \vec{H} - \tau_{(m)}
\frac{\partial}{\partial t}{\vec{M}} \,. \label{exa1}
\end{equation}
These equations are easily transformed into the general relaxation
equation
\begin{equation}
\frac{\partial}{\partial t}{\vec{\xi}} = - \frac{1}{\tau }
\left(\vec{\xi} - \vec{\xi}_{({\rm stationary})} \right)
\label{exa2}
\end{equation}
for the polarization-magnetization, see, e.g., \cite{LanKhal}. By
analogy with rheology, Kluitenberg \cite{Kluit1,Kluit2,Kluit3}
generalized the equations for polarization and magnetization
vectors of classical electrodynamics as
\begin{equation}
P^{\alpha} =
\chi^{\alpha}_{\beta} E^{\beta} - \sum^{(k)}_{(m)=(1)}
A^{\alpha}_{\beta (m)} \left(\frac{\partial}{\partial t}
\right)^{(m)} P^{\beta} +
\sum^{(s)}_{(m)=(1)} B^{\alpha}_{\beta (m)}
\left(\frac{\partial}{\partial t} \right)^{(m)} E^{\beta}
\label{kluit1} \,,
\end{equation}
\begin{equation}
M^{\alpha} = \zeta^{\alpha}_{\beta} H^{\beta} -
\sum^{(k)}_{(m)=(1)} C^{\alpha}_{\beta (m)}
\left(\frac{\partial}{\partial t} \right)^{(m)} M^{\beta} +
\sum^{(s)}_{(m)=(1)} D^{\alpha}_{\beta (m)}
\left(\frac{\partial}{\partial t} \right)^{(m)} H^{\beta} \,.
\label{kluit2}
\end{equation}
If the medium is anisotropic and there are  cross-effects, such as
pyro - electricity and pyro - magnetism, piezo - electricity and
piezo - magnetism \cite{Nye,Maugin88}, magneto - electricity
\cite{ODell}, electro - striction and magneto - striction
\cite{Nye,Maugin88}, etc., then the ``source"
$\chi^{\alpha}_{\beta} E^{\beta}$  in (\ref{kluit1}) and
$\zeta^{\alpha}_{\beta} H^{\beta}$ in (\ref{kluit2}) must be
supplemented by additional terms
\begin{equation}
\chi^{\alpha}_{\beta} E^{\beta}  \Rightarrow \chi^{\alpha}_{\beta}
E^{\beta} + \pi^{\alpha} (T-T_0) + d^{\alpha}_{\cdot \beta \gamma}
\sigma^{\beta \gamma} +
\nu^{\cdot \alpha}_{\beta} H^{\beta} + Q^{\alpha}_{\beta \gamma
\rho} E^{\beta} \sigma^{\gamma \rho} + ... \label{exa3}
\end{equation}
\begin{equation}
\zeta^{\alpha}_{\beta} H^{\beta}  \Rightarrow
\zeta^{\alpha}_{\beta} H^{\beta} + m^{\alpha} (T-T_0) +
h^{\alpha}_{\cdot \beta \gamma} \sigma^{\beta \gamma} +
\nu^{\alpha}_{\cdot \beta} E^{\beta} + R^{\alpha}_{\beta \gamma
\rho} H^{\beta} \sigma^{\gamma \rho} + ... \label{exa4}
\end{equation}
Here $\pi^{\alpha}$ and $m^{\alpha}$ are the pyro - electric and
pyro-magnetic coefficients, respectively, describing the variation
of polarization and magnetization produced by deviation of the
temperature from its equilibrium value, $T_0$. The coefficients
$d^{\alpha}_{\cdot \beta \gamma}$ and $h^{\alpha}_{\cdot \beta
\gamma}$ describe the linear piezo - electric  and piezo -
magnetic properties of the medium, respectively. These effects may
be induced by the stress tensor $\sigma^{\beta \gamma}$. The
tensor $\nu^{\alpha \beta}$ corresponds to magnetoelectric
coefficients; if they are non-vanishing, the medium transforms
electric fields into magnetic fields and vice-versa. The electro-
and the magneto - striction coefficients $Q^{\alpha}_{\beta \gamma
\rho}$ and $R^{\alpha}_{\beta \gamma \rho}$, describe
cross-effects involving the stress tensor and the electric or
magnetic field, respectively.

When the electrodynamic system as a whole is under motion, the
formulas (\ref{kluit1}) and (\ref{kluit2}) have to be supplemented
by the terms containing the acceleration vector and the spatial
derivatives of the velocity vector \cite{LLP}.

\subsubsection{Relativistic fluid in the absence of electromagnetic interactions}

The parabolic equations for the temperature evolution run into
conflict with the causality principle, since the rate of the
temperature propagation is predicted to be unbounded
\cite{Cattaneo58,Vernotte}. As an answer to this challenge, the
extended relativistic  (or causal, or second order, or transient)
thermodynamics was developed. The history, the fundamentals and
applications of  the extended thermodynamics are presented in
detail in Refs.  \cite{JCL,Muller} and
\cite{Israel2}-\cite{Herrera02}.

The extended covariant definition of heat flux four-vector:
\begin{equation}
q^i - \lambda T \Delta^i_k \left( \frac{1}{T} \nabla^k T - D U^k
\right) = \tau_{(1)} \Delta^{i}_k D q^k +
\frac{\tau_{(1)}}{2} q^i \left[ \Theta + D
\left(\log{\frac{\tau_{(1)}}{\lambda T^2}}\right) \right]
\label{dq}
\end{equation}
is a relativistic generalization of the definition
(\ref{cattaneo}). Here $D \equiv U^i \nabla_i$ is the convective
derivative, $\nabla_i$ is a covariant derivative operator; $\Theta
\equiv \nabla_k U^k$ is the fluid expansion. Both parts of the
relativistic tensor of non-Pascal pressure
\begin{equation}
\Pi^{ik} \equiv \Pi_{(0)}^{ik} + \frac{1}{3} \Delta^{ik} \Pi \,,
\quad \Pi \equiv g_{ik} \Pi^{ik} \label{PIik}
\end{equation}
obey expressions similar to (\ref{dq})
\begin{equation}
\Pi - 3 \zeta \Theta =  \tau_{(0)} D \Pi + \frac{\tau_{(0)}}{2}
\Pi \left[ \Theta + D \left(\log{\frac{\tau_{(0)}}{\zeta T}}
\right) \right] \,, \label{dPI}
\end{equation}
\begin{equation}
\Pi_{(0)}^{ik} -\eta \Sigma^{ik} = \tau_{(2)} \Delta^{i}_m
\Delta^{k}_n D \Pi_{(0)}^{mn} +
\frac{\tau_{(2)}}{2} \Pi_{(0)}^{ik} \left[ \Theta + D
\left(\log{\frac{\tau_{(2)}}{\eta T}}\right) \right] \,.
\label{dPI0}
\end{equation}
Here the well-known quantity $\Sigma^{ik}$ is introduced by
\begin{equation}
\Sigma^{ik} \equiv \frac{1}{2} \Delta^i_m \Delta^k_n (\nabla^m U^n
+ \nabla^n U^m) - \frac{1}{3} \Delta^{ik} \Theta \,. \label{Sigma}
\end{equation}
The principal novelty of the relativistic formulas (\ref{dq}) -
(\ref{dPI0}) is that they contain the convective derivative of the
velocity four-vector $DU^i$, the expansion scalar $\Theta$ as a
supplementary terms of the inertial origin.

\subsection{Remark on covariant electrodynamics of continuous media}

An obvious analogy exists between the constitutive equations in
electrodynamics and rheology. In this sense, $M^{ik}$ plays in
electrodynamics the role of a stress tensor $\sigma^{\alpha
\beta}$ in rheology, and $F^{ik}$ is an analog of the deformation
tensor $e_{\alpha \beta}$. The main difference is that while
electrodynamics deals with skew-symmetric quantities,
elastodynamics deals with symmetric ones. Following this analogy,
one can see a correspondence between the four-vector potential
$A_i$ in electromagnetism and  the displacement vector $V^{
\alpha}$ in classical elastodynamics. Indeed, the second subsystem
of Maxwell equations in (\ref{10maxwEQ}) leads to the relation
\begin{equation}
F_{ik} = \partial_i A_k - \partial_k A_i \,, \label{potential}
\end{equation}
which converts it into an identity. Analogously, from the
Saint-Venant conditions (equations) \cite{Maugin88} in classical
non-relativistic elastodynamics one obtains that the deformation
tensor has to be the symmetrized derivative of some three-vector,
the displacement vector $V^{\alpha}$:
\begin{equation}
e_{\alpha \beta} = \frac{1}{2} (\partial_{\alpha} V_{\beta} +
\partial_{\beta} V_{\alpha}) \,. \label{saintvenant}
\end{equation}
Likewise, there is an analogy between the constitutive laws in the
electrodynamics and elastodynamics. The well-known linear static
constitutive equations in electrodynamics says that $H^{ik}$ is
proportional to $F_{mn}$
\begin{equation}
H^{ik} = C^{ikjl} F_{jl} \equiv H^{ik}_{({\rm stationary})}  \,,
\label{HCF}
\end{equation}
where the tensor $C^{ikjl}$ is the linear response tensor. In
contrast to (\ref{hook}) with symmetric tensor ${\cal C}^{\alpha
\beta \gamma \rho}$ (\ref{elacoeff}) the relations (\ref{HCF})
contain the tensor $C^{ikjl}$ (\ref{eldacoeff}) with
skew-symmetric indices in the first and the second pairs.
Alternatively, the polarization - magnetization tensor $M^{ik}$ is
considered  proportional to the Maxwell tensor
\begin{equation}
M^{ik} = \chi^{ikmn} F_{mn} \equiv M^{ik}_{({\rm stationary})} \,.
\label{MchiF}
\end{equation}
Here $\chi^{ikmn}$ is called the susceptibility tensor, it has the
same symmetry of indices, as $C^{ikmn}$. Relation (\ref{MchiF})
is, in fact, the analog of the Hook expression (\ref{hook}). The
covariant phenomenological generalization of the constitutive
equations may also be done in terms of polarization and
magnetization four-vectors $P^i$ and $M^i$:
\begin{equation}
P^i \equiv M^{ik}U_k \,, \quad M_i \equiv M^{*}_{ik}U^k \,.
\label{PMUMMU}
\end{equation}
The equations (\ref{MchiF}) yield
\begin{equation}
P^i = \alpha^{ik} E_k - \gamma^{ \ i}_{k \cdot} B^k \,, \quad M_i
= \gamma^{ \ k}_{i \cdot } E_k + \beta_{ik} B^k \,, \label{PEBMEB}
\end{equation}
where $\alpha^{ik}$, $\beta_{ik}$ and $\gamma^{ \ i}_{k \cdot}$
can be obtained from (\ref{emunu}) by the substitution $C^{ikmn}
\Rightarrow \chi^{ikmn}$. Relativistic covariant elastodynamics is
much more sophisticated (see, e.g.,
\cite{Tsypkin,Chernyi,Beig1,Beig2} and references therein), and
the described analogy is not so evident. Nevertheless, let us use
the main idea of generalization of the elastodynamic constitutive
equations in order to obtain extended constitutive equations for
the covariant electrodynamics.

\subsection{ Phenomenologically extended constitutive equations for
relativistic electrodynamic systems}

Based on the analogies described in the previous Section we now
introduce generalized phenomenological constitutive equations for
non- stationary electromagnetic media. We consider three versions
of generalization, which, of course, are equivalent, but can be
useful for different applications.

\subsubsection{The first version}

The first version of the extended electrodynamics of continuous
media  assumes that the difference between $M^{ik}$  and its
stationary value $\chi^{ikjl} F_{jl}$ can be written as
\begin{equation}
M^{ik} -  \chi^{ikjl} F_{jl}  {=} \Gamma^{ik}_{\cdot \cdot mn} D
M^{mn} {+} \Lambda^{ik}_{\cdot \cdot mn} D F^{mn} {+}
\Omega^{ikm} DU_m {+} ... ({\rm higher} \ {\rm derivative} \
{\rm terms})\,. \label{Iversion}
\end{equation}
Here the velocity four-vector enters explicitly the convective
derivative $D= U^k \nabla_k$ only. It is clear that the structure
of the formula (\ref{Iversion}) is analogous to the
Poynting-Thomson law (\ref{thomson}).

\subsubsection{The second version}

More generally, one can replace the convective derivative, $D$, by
the covariant derivative;  then we obtain
\begin{equation}
M^{ik} {-} \chi^{ikjl} F_{jl} {=} \Gamma^{ikj}_{\cdot \cdot \cdot
mn} \nabla_j M^{mn} {+} \Lambda^{ikj}_{\cdot \cdot \cdot mn}
\nabla_j F^{mn} {+}
\Omega^{ikmj} \nabla_j U_m {+} ({\rm higher} \ {\rm derivative}
\ {\rm terms}). \label{IIversion}
\end{equation}
In the last case it is convenient to use the standard
decomposition
\begin{equation}
\nabla_j U_l = U_j D U_l + \Sigma_{jl} + \frac{1}{3} \Delta_{jl}
\Theta + \omega_{jl} \,. \label{nablaU}
\end{equation}
The shear tensor $\Sigma_{jm}$ was defined in (\ref{Sigma}), the
scalar expansion $\Theta$ is again $\nabla_k U^k$, and the
vorticity tensor $\omega_{jl}$ is defined by
\begin{equation}
\omega_{jl} = \frac{1}{2} \Delta_j^{m} \Delta_l^{n} (\nabla_m U_n
- \nabla_n U_m) \,. \label{omega}
\end{equation}
Obviously, equation (\ref{IIversion}) reduces to (\ref{Iversion})
when
\begin{equation}
\Gamma^{ikj}_{\cdot \cdot \cdot mn} = U^j \Gamma^{ik}_{\cdot \cdot
mn}  \,, \quad \Lambda^{ikj}_{\cdot \cdot \cdot mn} = U^j
\Lambda^{ik}_{\cdot \cdot mn} \,,
\quad \Omega^{ikmj} = U^j \Omega^{ikm} \,. \label{link}
\end{equation}
In general there exist  standard irreducible decompositions of
such tensors similar to the decomposition of the $C^{ikmn}$ tensor
(\ref{Cdecomp}). The constitutive law (\ref{IIversion}) is
applicable not only to the non-stationary models, but to the
non-homogeneous media, as well. In the latter case the difference
$M^{ik} - M^{ik}_{({\rm stationary})}$ should be decomposed using
the spatial derivatives, $\Delta^{k}_i \nabla_k$, in addition to
convective derivative $D= U^k \nabla_k$. Moreover, the presence of
higher derivative terms in (\ref{IIversion}) gives the possibility
to consider cross-terms of the type (\ref{cattaneoG}), described
in \cite{Jou1} for heat conduction.

\subsubsection{The third version}

Consider now the tetrad $\{U^i, X^i_{(\alpha)} \}$, where
$(\alpha) = (1),(2),(3)$. Here $X^i_{(0)} \equiv U^i$ is the
velocity four-vector of the medium and the tetrad vectors
$X^i_{(\alpha)}$ are connected with three  main directions in the
anisotropic medium. Define the tetrad components of the vectors of
polarization and magnetization,  of the vectors of electric field
and magnetic induction, as well the acceleration vector as
follows:
\begin{equation}
P_{(\alpha)} \equiv P_i X^i_{(\alpha)} \,, \quad M_{(\alpha)}
\equiv M_i X^i_{(\alpha)} \,, \quad  E_{(\alpha)} \equiv E_i
X^i_{(\alpha)}\,,
\quad B_{(\alpha)} \equiv B_i X^i_{(\alpha)} \,, \quad (DU)_{(\alpha)}
\equiv X^i_{(\alpha)} \ DU_i \,. \label{s2}
\end{equation}
Since the vectors $P^i$, $M^i$, $E^i$, $B^i$ and $DU^i$ are
orthogonal to the velocity four-vector and the vectors
$X^i_{(\alpha)}$ are space-like, we have the inverse
decompositions in the form:
\begin{equation}
P^i  {=} {-} \sum_{(\alpha)} P_{(\alpha)} X^i_{(\alpha)} \,, \quad
M^i  {=} {-} \sum_{(\alpha)} M_{(\alpha)} X^i_{(\alpha)} \,, ...
\,. \label{s3}
\end{equation}
As a main ansatz for the third version we suggest that in the
appropriate tetrad the relaxation equations take the form
\begin{equation}
P^{(\alpha)} = \alpha^{(\alpha)}_{(\beta)} E^{(\beta)} - \gamma^{
\ (\alpha)}_{(\beta) \cdot } B^{(\beta)} -
\tau^{(\alpha)}_{(\beta)({\rm p})} D P^{(\beta)} +
\lambda^{(\alpha)}_{(\beta)} D E^{(\beta)} +
\xi^{(\alpha)}_{(\beta)} D B^{(\beta)} + \eta^{(\alpha)}_{(\beta)}
(DU)^{(\beta)} + ...\,, \label{IIIversion1}
\end{equation}
\begin{equation}
M^{(\alpha)} = \gamma^{ \ (\alpha)}_{(\beta) \cdot} E^{(\beta)} +
\beta^{(\alpha)}_{(\beta)} B^{(\beta)} -
\tau^{(\alpha)}_{(\beta)({\rm m})} D M^{(\beta)} + \kappa^{(\alpha)}_{(\beta)} D E^{(\beta)} +
\psi^{(\alpha)}_{(\beta)} D B^{(\beta)} +
\zeta^{(\alpha)}_{(\beta)} (DU)^{(\beta)} +...\,.
\label{IIIversion2}
\end{equation}
In such a context we assume, that the Einstein rule for the
indices $(\beta)$ are valid. The term
$\tau^{(\alpha)}_{(\beta)({\rm p})}$ describes the diagonal
three-dimensional matrix of the relaxation coefficients for the
corresponding tetrad components of the polarization vector.
Analogously, three independent coefficients of the diagonal matrix
$\tau^{(\alpha)}_{(\beta)({\rm m})}$ describe three relaxation
parameters, which differ in general for three tetrad components of
the magnetization vector. We assume that different components of
the polarization and magnetization vectors evolve with its own
relaxation time parameters. The constitutive equations
(\ref{IIIversion1}) and (\ref{IIIversion2}) generalize the
formulas (\ref{kluit1}) and (\ref{kluit2}), advocated by
Kluitenberg \cite{Kluit1,Kluit2}.

The first and the third versions of the generalization are
equivalent, when the law of the tetrad vectors evolution is fixed.
To satisfy the conditions of orthogonality-normalization for the
tetrad vectors, one can use, e.g., the simplest expression
\begin{equation}
D X^i_{(\alpha)} = \Omega^{i}_{ \ k}  \ X^k_{(\alpha)} \,,
\label{DX}
\end{equation}
where the tensor $\Omega_{ik}$ is skew-symmetric. Then using
\begin{equation}
M^{ik} = P^i U^k - P^k U^i - \eta^{ikj} M_j \,, \label{MPM}
\end{equation}
with (\ref{s3}) and (\ref{DX}), and expressing the coefficients
$\Gamma^{ik}_{\cdot \cdot mn}$, ..., etc.,  via the tetrad
coefficients $\tau^{(\alpha)}_{(\beta)({\rm p})}$,
$\tau^{(\alpha)}_{(\beta)({\rm m})}$  ...,  etc., one obtains the
third version (\ref{IIIversion1}),(\ref{IIIversion2}) of the
generalized constitutive law from the first one (\ref{Iversion}).

\subsubsection{On the electrodynamics of  thermo-visco-elastic medium}

When thermo-visco-elastic processes take place, the
right-hand-sides of the constitutive equations (\ref{Iversion}),
 (\ref{IIversion}), (\ref{IIIversion1}), (\ref{IIIversion2}) have
to be supplemented by the heat-flux vector $q^i$ and its
derivatives $D^{(m)}q^i$, the non-equilibrium pressure $\Pi^{ik}$
and its derivatives $D^{(m)}\Pi^{ik}$ as discussed  above. In its
turn, the constitutive equations for $q^i$ and $\Pi^{ik}$ have to
be supplemented by the corresponding electromagnetic terms
$M^{ik}$, $F_{mn}$,... $D^{(m)}M^{ik}$, $D^{(m)}F_{mn}$, etc. This
problem, of course, requires a special consideration which takes
into account the concept of hyperbolicity of master equations
(see, e.g., \cite{Beig3}). Thus, in general, one obtains a system
of coupled extended constitutive equations for $M^{ik}$, $q^i$ and
$\Pi^{ik}$, describing the covariant extended electrodynamics of
continua.

\section{Example of exact solution of non - stationary Einstein - Maxwell model}

\subsection{Einstein's equations}

We consider the FLRW cosmological model with line element
\cite{ExactSolutions,MTW}
\begin{equation}
ds^2 = dt^2 - a^2(t) \ \left[(dx^1)^2 + (dx^2)^2 + (dx^3)^2
\right] \,. \label{metric}
\end{equation}
The magnetic field vector, the  magnetization vector,  the
electric field vector and the polarization vector are pointed
along the $x^3$ axis. For such a ``self-parallel" configuration of
the electromagnetic field the total stress-energy tensor
$T^{ik}_{({\rm total})}$, has four non-vanishing components:
\begin{equation}
T^0_{0({\rm total})} {=} W {+} X \,, \quad T^1_{1({\rm total})}
{=} {-} P_{(1)} {-} X \,,
\quad T^2_{2({\rm total})} {=} {-} P_{(2)} {-} X \,, \quad T^3_{3({\rm
total})} {=} {-} P_{(3)} {+} X \,. \label{Teff1}
\end{equation}
Here
\begin{equation}
X \equiv \frac{1}{2} [H^{12} F_{12} - H^{30} F_{30}]  \label{Xdef}
\end{equation}
is the source term related to the electric and magnetic fields,
polarization  and magnetization. We do not specify here the
relations between the energy density scalar $W$ and the diagonal
components of the pressure tensor $P_{(1)}$, $P_{(2)}$ and
$P_{(3)}$. They can describe perfect fluid, viscous fluid, etc. In
a co-moving frame with $U^i=\delta^i_0$, the gravity field
equations reduce to the following system \cite{MTW}
\begin{equation}
3 \left(
\frac{\dot{a}}{a} \right)^2 = \Lambda + \kappa ( W  + X) \,,
\quad
2 \frac{\ddot{a}}{a} + \left( \frac{\dot{a}}{a} \right)^2 =
\Lambda - \kappa ( P_{(1)} + X )  \,, \label{Ein1}
\end{equation}
\begin{equation}
2\frac{\ddot{a}}{a} + \left( \frac{\dot{a}}{a} \right)^2 = \Lambda
- \kappa ( P_{(2)} + X ) \,,
\quad
2 \frac{\ddot{a}}{a} + \left( \frac{\dot{a}}{a} \right)^2 =
\Lambda - \kappa ( P_{(3)} - X ) \,. \label{Ein2}
\end{equation}
The dot denotes the derivative with respect to time. The Einstein
equations are self-consistent when
\begin{equation}
P_{(1)} + X = P_{(2)} + X = P_{(3)} - X = {\cal P}_{({\rm
isotr})}\,. \label{self}
\end{equation}
The formulas (\ref{self}) guarantee that the global spatial
isotropy of the Universe holds. Differentiating (\ref{Ein1}) -
(\ref{Ein2}) leads to the conservation law
\begin{equation}
\dot{W} {+} \dot{X} {+} \left(\frac{\dot{a}}{a} \right)
\left[4X{+}3W {+} P_{(1)} {+} P_{(2)} {+} P_{(3)} \right] {=} 0 .
\label{conserva}
\end{equation}
In principle, the matter pressure may be anisotropic, but with
$P_{(1)} = P_{(2)} = P_{(3)} - 2X$, i.e., the longitudinal
pressure $P_{(3)}$ compensates the influence of the
electromagnetic pressure. However, the question arises: whether a
non-trivial solution exists for which electromagnetic field and
electromagnetic induction are non-vanishing, but the matter has an
isotropic pressure $P_{(1)}= P_{(2)}= P_{(3)} = P$.  Obviously,
such a requirement  assumes, that  $X=0$. For the well-known
models with magnetic field \cite{dina} this requirement leads
automatically to the absence of the magnetic field. Is it possible
to self-consistently consider a {\it non-trivial} electromagnetic
field in the FLRW background? The answer is yes. Consider such a
model.

Given the structure of Einstein's field equations, in the model of
parallel electric and magnetic fields the following
electromagnetic source governs the evolution of the gravitational
field:
$$
X \equiv \frac{1}{2}(H^{12}F_{12} - H^{30}F_{30})
= \frac{1}{2}(M^{12}F_{12} +  F^{12}F_{12} - M^{30}F_{30} -
F^{30}F_{30})
$$
\begin{equation}
= - \frac{1}{2}\left(  M^{3}B_{3} + B^{3}B_{3} + P^{3}E_{3} +
E^{3}E_{3} \right)
= \frac{1}{2}\left( M^{(3)}B^{(3)} {+} (B^{(3)})^2 {+}
P^{(3)}E^{(3)} {+} (E^{(3))^2} \right) \,. \label{X}
\end{equation}
When both the polarization vector $P^{(3)}$ and magnetization
vector $M^{(3)}$ vanish, then the quantity $X(t)$ is non-negative.
Nevertheless, when $P^{(3)}$ and $M^{(3)}$, are non-vanishing,
$X(t)$ can be negative for some time interval, or even identically
vanish for the special initial conditions. Note that even if $X$
is negative, the total energy density $W+X$ is assumed to be
non-negative, since the energy density of the material medium,
$W$, is positive. We consider below just the case with $X=0$.

\subsection{Maxwell's equations}

The quantities $F_{ik}$ and $M_{ik}$ are considered to be the
function of cosmological time only. Thus, it follows from the
second subsystem of Maxwell equations that the tensor of
electromagnetic field has a few constant components, and using the
symmetries of the model, we can choose only $F_{12} = const$ to be
non-vanishing, since the magnetic field points along the $x^3$
axis. Note that the equations (\ref{10maxwEQ}) do not fix the
structure of the  electric field, so far an arbitrary function of
time. We assume for the simplicity that the current $I^i$
vanishes. The first subsystem of the Maxwell equations
\begin{equation}
\nabla_k H^{ik} = \frac{1}{\sqrt{{-}g}} \partial_k (\sqrt{{-}g}
H^{ik}) = \frac{1}{a^3}\frac{d}{dt} (a^3 H^{i0}) = 0
\label{maxwell1}
\end{equation}
reduces to an identity for $i=0$ and three differential equations
in ordinary derivatives for the three components $H^{\alpha 0}$,
whose solution is
\begin{equation}
H^{\alpha 0}(t) \equiv M^{\alpha 0}(t) + F^{\alpha 0}(t) =
P^{\alpha}(t) + E^{\alpha }(t) =
\frac{const}{a^3(t)} = D^{\alpha}(t_0) \frac{a^3(t_0)}{a^3(t)}\,.
\label{Hconst}
\end{equation}
Note that the last relation includes the polarization vector
$P^{\alpha}(t)$ and electric field $E^{\alpha }(t)$ only. Thus,
the evolution of the magnetization vector $M^{\alpha}$ is just
governed by the constitutive equations.

\subsection{Constitutive equations}

We use  the third version of the generalized constitutive
equations (\ref{IIIversion1}) and (\ref{IIIversion2}). For this
case, we specify the tetrad vectors. For the metric (\ref{metric})
the normalization - orthogonality conditions for the tetrad
vectors yield
\begin{equation}
U^i = \delta^i_0 \,, \quad X^i_{(\alpha)} = \frac{1}{a(t)}
\delta^i_{\alpha}\,. \label{tetrada}
\end{equation}
For such a four-velocity vector the acceleration vector $DU^i$
vanishes. Consider the simplest relaxation model without electric
conductivity, in which the polarization is coupled to magnetic
field by the non-vanishing magnetoelectric coefficients. The
master equations for such a model read
\begin{equation}
\tau_{({\rm p})} \dot{{\cal P}} + {\cal P} = (\varepsilon_{||} -1)
E - \gamma B \,, \label{5dP}
\end{equation}
\begin{equation}
\tau_{({\rm m})} \dot{M} + M =  \gamma E +
\left(\frac{1}{\mu_{||}} -1 \right) B  \,, \label{5dM}
\end{equation}
\begin{equation}
E(t) + {\cal P}(t) = D(t_0) \left( \frac{a(t_0)}{a(t)} \right)^2
\,, \label{5DEP}
\end{equation}
\begin{equation}
B(t) = B(t_0) \left( \frac{a(t_0)}{a(t)} \right)^2 \,, \quad
B(t_0) = \frac{F_{12}}{a^2(t_0)} \,. \label{5BF}
\end{equation}
Here we use the quantities ${\cal P} \equiv P^{(3)}$, $E \equiv
E^{(3)}$, $B \equiv B^{(3)}$, $D \equiv D^{(3)}$, ... etc., in
which, for simplicity, we omit the tetrad indices. Assuming that
the relaxation parameters depend on time according to
\begin{equation}
\tau_{({\rm p})} = \xi_1 \left(\frac{\dot{a}}{a} \right)^{-1}\,,
\quad \tau_{({\rm m})} = \xi_2 \left(\frac{\dot{a}}{a}
\right)^{-1} \,, \label{5xi12}
\end{equation}
($\xi_1 , \xi_2 = {\rm const}$) and introducing the variable $x =
\frac{a(t)}{a(t_0)}$, we obtain that the solutions to (\ref{5dP})
- (\ref{5BF}) are
\begin{equation}
{\cal P}(x) = {\cal P}(t_0) x^{- \varepsilon_{||} / \xi_1} +
\Gamma_1 \left( x^{-2} -  x^{- \varepsilon_{||} / \xi_1} \right)
\,, \label{5P}
\end{equation}
\begin{equation}
M(x) = M(t_0) x^{- \frac{1}{\xi_2}} + \Gamma_2 \left( x^{-2} -
x^{- \frac{1}{\xi_2}} \right) +
\Gamma_3 \left( x^{- \varepsilon_{||} / \xi_1} - x^{-
\frac{1}{\xi_2}} \right) \,, \label{5M}
\end{equation}
where
\begin{equation}
\Gamma_1 \equiv \left[ \frac{D(t_0) (\varepsilon_{||} -1) - \gamma
B(t_0)}{(\varepsilon_{||} -2 \xi_1)} \right] \,, \label{5Gamma1}
\end{equation}
\begin{equation}
\Gamma_2  \equiv \frac{1}{(1 - 2 \xi_2)(\varepsilon_{||} - 2
\xi_1)} \left\{ B(t_0) \left[\gamma^2 + \left(\frac{1}{\mu_{||}} {-}1
\right)(\varepsilon_{||} {-} 2 \xi_1) \right] {+} \gamma D(t_0) (1
{-} 2 \xi_1)\right\} \,, \label{5Gamma21}
\end{equation}
\begin{equation}
\Gamma_3  \equiv \frac{\gamma \xi_1}{(\xi_1 - \xi_2
\varepsilon_{||})} \left[ \Gamma_1 - {\cal P}(t_0) \right] \,.
\label{5Gamma3}
\end{equation}
Thus, the function $X(t)$ reads
$$
X(t) {=} \frac{1}{2} x^{(-2 - \frac{1}{\xi_2})} B(t_0) [M(t_0) {-}
\Gamma_2 {-} \Gamma_3 ] {+}
\frac{1}{2} x^{- 4} [B^2(t_0) {+} D^2(t_0) {+} B(t_0) \Gamma_2
{-} D(t_0) \Gamma_1]
$$
\begin{equation}
{+} \frac{1}{2} x^{({-} 2{-} \varepsilon_{||} / \xi_1)} [B(t_0)
\Gamma_3 {+} D(t_0) \Gamma_1 {-} D(t_0) {\cal P}(t_0)] \,.
\label{5X}
\end{equation}
Since $X(t)$ contains nine free parameters, we can choose three of
them, for instance, the initial data, ${\cal P}(t_0)$, $M(t_0)$
and $D(t_0)$, so that $X(t)=0$. In other words, we have a model
with ``hidden" electric and magnetic fields. For instance, when
\begin{equation}
{\cal P}(t_0) = \Gamma_1 \,, \quad M(t_0) = \Gamma_2 \,,
\label{X000}
\end{equation}
it follows that
\begin{equation}
{\cal P}(x) {=} \Gamma_1 \ x^{{-}2} \,, \quad E(x) {=} [D(t_0) {-}
\Gamma_1] \ x^{{-}2} \,,
\quad M(x) {=} \Gamma_2 \ x^{{-}2} \,, \quad H(x) {=} [B(t_0) {+}
\Gamma_2] \ x^{{-}2} \,. \label{mp000}
\end{equation}
Thus, $X=0$ when
\begin{equation}
B^2(t_0) + D^2(t_0) + B(t_0) \Gamma_2 -  D(t_0) \Gamma_1 = 0 \,.
\label{000}
\end{equation}
This condition reduces to the quadratic equation for the ratio
$D(t_0) / B(t_0)$.
\begin{equation}
\left(\frac{D(t_0)}{B(t_0)} \right)^2 (1 {-} 2 \xi_1)(1 {-} 2
\xi_2) {+} 2 \gamma \frac{D(t_0)}{B(t_0)} (1 {-} \xi_1 {-} \xi_2)
{+}\left[ \gamma^2 {+} \left( \frac{1}{\mu_{||}} {-} 2 \xi_2
\right) \left( \varepsilon_{||} {-} 2 \xi_1 \right) \right] = 0
\,. \label{quadra}
\end{equation}
This equation has real roots for a wide choice of the parameters
$\xi_1$, $\xi_2$, $\gamma$, $\mu_{||}$ and $\varepsilon_{||}$. For
instance, when $\xi_1 = \xi_2 = \xi$ and $\gamma=0$ two real
solutions exist when $\frac{1}{\mu_{||}} < 2 \xi <
\varepsilon_{||}$, i.e.,  when the relaxation parameters
$\tau_{({\rm p})}$ and  $\tau_{({\rm m})}$ are of the order of
Hubble parameter $H(t) {=} \dot{a} / a$.

Thus, we obtained an exact solution of the Einstein-Maxwell
equations, describing the FLRW-type model, in which there is a
non-vanishing magnetic field, the magnetization, the electric
field and the polarization of the matter, however, they are
hidden, i.e., their total contribution to the stress-energy tensor
of the whole system vanishes. This type of behaviour was discussed
in \cite{BZ05}. There the stationary magnetic field in vacuum is
non-vanishing, nevertheless, the exact solution to the non-minimal
Einstein-Maxwell equations demonstrates the possibility of
isotropic FLRW-type expansion. In that case the non-minimal
interaction between gravitational and electromagnetic fields
inspires some kind of ``non-minimal screening" and hiddens
magnetic field from the point of view of gravitational dynamics.
Here we presented the example of ``dynamic screening", when the
polarization and magnetization of the medium compensate the
contribution of the electromagnetic field to the total
stress-energy tensor. Note that in the proposed model the magnetic
induction, $H^{(3)} = B^{(3)} + M^{(3)}$, and the electric
induction, $D^{(3)} = E^{(3)} + P^{(3)}$, are considered to be
non-vanishing. This means, that the electric polarization
compensates (partially) the contribution of the magnetic field to
the expression for $X$, and the  magnetization compensates
(partially) the contribution of the electric field to $X$ because
of the special choice of the initial data. When the electric field
and electric polarization are absent, there exists, nevertheless,
the possibility that the magnetization compensates the
contribution of the magnetic field. It is possible due to the
relationship $2 \mu_{||} \xi_2 = 1$. The magnetic induction is
vanishing in this case.

When $X{=}0$ the Universe is expanding isotropically and does not
feel the presence of the electric and magnetic fields. The
Einstein equations for this case are the standard (see,
(\ref{Ein1}) and (\ref{Ein2})  with $X{=}0$ and
$P_{(1)}{=}P_{(2)}{=}P_{(3)}{=}P$). We will not specify the
equation of state and discuss the solutions. Nevertheless, let us
note, that the application of the extended constitutive equations
to the cosmic electrodynamics have something in common with
inhomogeneous (depending on time) equations of state, introduced
in \cite{Odintsov1} - \cite{Odintsov4}.

\section{Conclusions}

The fundamentals of covariant phenomenological electrodynamics of
relativistic continuous media were elaborated three decades ago,
however, the self-consistent description of a gravitating
polarizable - magnetizable non - stationary medium is still an
open question. In this paper we formulated extended Einstein -
Maxwell model appropriated for the non-stationary
electromagnetically active relativistic material medium, which has
three main ingredients:

{\it (i)}. The standard Maxwell equations, describing the
electrodynamic phenomena in continuous media in terms of induction
tensor and Maxwell tensor (see (\ref{0maxwEQ}) and
(\ref{10maxwEQ})).

{\it (ii)}. A new set of covariant constitutive equations,
containing the polarization - magnetization tensor and its first,
second, etc., covariant derivatives (see (\ref{Iversion}),
(\ref{IIversion}), (\ref{IIIversion1}) and (\ref{IIIversion2})).
In this sense the model can be indicated as an {\it extended} one
in analogy with extended thermodynamics.

{\it (iii)}.  The Einstein equations for the gravity field, in
which we introduced a new effective stress-energy tensor
(\ref{BPSET}), describing the contribution of the electromagnetic
field, of the electric polarization and of the magnetization of
the medium. This Einstein-Maxwell model can be indicated as {\it
self-consistent} by two reasons. First, the corresponding extended
constitutive equations describe the interaction of matter with
electromagnetic field, resulting in the dynamics of polarization -
magnetization. Secondly, the polarization and  magnetization
contribute to the total stress - energy tensor, the source for the
gravitational field,  via the proposed effective stress - energy
tensor of the electromagnetic field.

Concerning our approach, we would like to emphasize three points.

a) This extended model needs verification. In this paper we
discuss only one example of the exact solution to the extended
Einstein-Maxwell model, describing the FLRW-type cosmological
dynamics with hidden electromagnetic field. We prepared also a
paper related to Bianchi-I anisotropic cosmological model with
exact solutions of the new type. We hope the extended
Einstein-Maxwell model will be also useful in application to the
theory of interaction of the gravitational waves with
electromagnetically active material media.

b) This model admits also a generalization to gravitating static
anisotropic {\it non-homogene\-ous} media. To develop such a model
one can introduce the first, second, etc. {\it spatial}
derivatives of the polarization-magnetization tensor into the
constitutive equations in analogy to the convective derivative.

c) We introduced the extended constitutive equations by the
phenomenological way. The next step is to confirm this approach by
the consideration of the corresponding entropy production scalar
and by analyzing the first and the second laws of thermodynamics.
We shall present such analyses in future papers.

\vspace{8mm}
\noindent
{\bf Acknowledgements}

\noindent
This work was supported by the Catalonian Government
(grant 2004 PIV1 35) and was done in the Department of Physics of
the Universidad Aut\'onoma de Barcelona. The author is grateful to
the colleagues of Group of Statistical Physics for hospitality.
The author is especially thankful to Prof. D. Pav\'on for reading
the manuscript, helpful discussions and advices.

\end{document}